\documentstyle[12pt]{article}

\begin{document}

\begin{center}
THE TWO-DIMENSIONAL ANALOGUE OF GENERAL RELATIVITY
\vskip 1cm
{\bf Jos\'e P. S. Lemos} \\
\vskip 0.3cm
{\scriptsize  Departamento de Astrof\'{\i}sica,
              Observat\' orio Nacional-CNPq,} \\
{\scriptsize  Rua General Jos\'e Cristino 77,
              20921 Rio de Janeiro, Brazil,} \\
{\scriptsize  \&} \\
{\scriptsize  Departamento de F\'{\i}sica,
              Instituto Superior T\'ecnico,} \\
{\scriptsize  Av. Rovisco Pais 1, 1096 Lisboa, Portugal.} \\

\vskip 0.6cm
{\bf Paulo M. S\'a} \\
\vskip 0.3cm
{\scriptsize  Centro de F\'{\i}sica da Mat\'eria Condensada,} \\
{\scriptsize  Av. Prof. Gama Pinto 2,
              1699 Lisboa Codex, Portugal.}
\end{center}

\bigskip

\begin{abstract}
\noindent
General Relativity in three or more dimensions can be obtained
by taking the limit $\omega\rightarrow\infty$ in the
Brans-Dicke theory.
In two dimensions General Relativity is an unacceptable theory.
We show that the two-dimensional closest analogue
of General Relativity is a theory that also arises in
the limit $\omega\rightarrow\infty$ of the two-dimensional
Brans-Dicke theory.
\end{abstract}

\newpage

\noindent
{\bf 1. Introduction}
\vskip 3mm
\noindent
It is known that in two-dimensional (2D) spacetimes the
Einstein-Hilbert action,
$S=\int d^2x \sqrt{-g}R$,
is a topological invariant;
variation with respect to the metric $g_{ab}$ yields the
Einstein tensor,
$G_{ab} \equiv R_{ab}-\frac12 g_{ab} R$,
which is identically zero for any 2D metric.
The energy-momentum tensor is defined as
\begin{equation}
T_{ab}\equiv\frac{2}{\sqrt{-g}}
  \frac{\delta (\sqrt{-g} {\cal L}_{matter})}{\delta g_{ab}},
\label{eq:1}
\end{equation}
where ${\cal L}_{matter}$ is the Lagrangean for the matter
fields.
Then, assuming Einstein' s equations,
$G_{ab}-\Lambda g_{ab}=T_{ab}$,
the vanishing of $G_{ab}$ implies that
$T_{ab}$ should be proportional to $\Lambda g_{ab}$,
giving an unacceptable theory.
Thus, one has to consider alternative
2D theories of gravitation.

Teitelboim and Jackiw \cite{teit} have proposed the simplest model
for a 2D theory of gravitation,
given by the constant curvature equation
$R-\Lambda=0$.
Introducing an auxiliary scalar field, $\Phi(x)$,
one can derive this equation through an action
given by
\begin{equation}
   S=\int d^2x \sqrt{-g} \Phi(x) \left( R-\Lambda \right).
\label{eq:2}
\end{equation}
To include matter one can add the trace $T$ of an
energy-momentum tensor $T_{ab}$ to yield the equation
$R-\Lambda=T$ \cite{brow,mann1}.
However, to derive this equation from a variational principle
it is now difficult to justify the inclusion of the trace
$T$ inside the action of eq.~(\ref{eq:2}).
To remedy this problem a different action with a different
auxiliary field $\psi$ has been proposed
\cite{torr,mann2}
\begin{equation}
   S=\int d^2x \sqrt{-g}
     \left(  \psi R
            +\frac12 (\partial \psi)^2
            +\Lambda
            +2{\cal L}_{matter} \right).
\label{eq:3}
\end{equation}
This action gives the desired equation
$R-\Lambda=T$, where $T$ is the trace
of the energy-momentum tensor defined in
eq.~(\ref{eq:1}).

Now, a general (not the most general) 2D
action can be written as
\begin{equation}
   S=\int d^2x \sqrt{-g} e^{-2\phi}
     \left( R-4\omega(\partial\phi)^2+4\lambda^2 \right),
\label{eq:4}
\end{equation}
where $\omega$ is a parameter,
$\lambda$  is a constant usually related to $\Lambda$ as
$\Lambda=-4\lambda^2$, and
$\Phi$ and $\phi$ are related through $\Phi=e^{-2\phi}$.
This action includes several important cases:
for $\omega=0$ one recovers the Teitelboim-Jackiw theory
given in eq.~(\ref{eq:2}),
$\omega=-\frac12$ gives planar General Relativity
\cite{lemo1}
and for $\omega=-1$ one obtains the first
order string theory \cite{mand,witt}.
This 2D Brans-Dicke theory has been analised by
the authors \cite{sa} for all values of $\omega$ and
of the cosmological constant $\lambda$.
It admits various types of black holes with different
types of singularities.

One question that is still open, is which of the 2D theories
of gravity is the analogue of General Relativity.
In $n$-dimensions ($n\geq 3$) the Einstein's theory of gravity
can be obtained from the Brans-Dicke theory by taking in the
latter the limit $\omega\rightarrow\infty$.
In particular, this holds for our four-dimensional world.
So, it is natural to consider that the analogue of
General Relativity in two-dimensions is also the theory
of gravity obtained from 2D Brans-Dicke theory in the
abovementioned limit.
In this letter we show that the theory given by action
(\ref{eq:3}) is,
in that sense,
the 2D analogue of Einstein's General Relativity.
Indeed, in the limit $\omega\rightarrow\infty$ the
2D Brans-Dicke theory (\ref{eq:4})
(generalized to include a matter Lagrangean) gives rise to the
equation $R-\Lambda=T$.
This theory has been called the $R=T$ theory \cite{mann3}.

In section~2 we perform a heuristic derivation
of the 2D Brans-Dicke action,
and in section~3 we show how in the limit
$\omega\rightarrow\infty$ of Brans-Dicke theory
one obtains the 2D analogue of General Relativity.
Finally we summarize our results
in a concluding section.

\vskip 1cm
\noindent
{\bf 2. An Heuristic Derivation of the 2D Brans-Dicke Action}
\vskip 3mm
\noindent
Let us start with Newtonian dynamics in order
to arrive at a relativistic theory.
In one spatial dimension,
Laplace's equation implies that the force $F$
between particles on a line is constant.
This in turn implies that there are no tidal forces on a line,
$dF=0$.
Now, infering that
the 2D world is also special relativistic,
one has to generalize the 2D Newtonian theory.
If one associates tidal forces with the curvature,
then Newton's law implies that the scalar curvature
vanishes, $R=0$.
A theory with this equation cannot be derived from the
Hilbert action, $S=\int d^2x \sqrt{-g} R$,
since it gives
$G_{ab}=R_{ab}-\frac12 g_{ab}R\equiv0$,
which is always true in 2D.
However,
it can be derived from
$S=\int d^2x \sqrt{-g} e^{-2\phi }R$,
where $\phi$ is a scalar field.
Variation with respect to $\phi$,
$\frac{\delta S}{\delta\phi}$, yields $R=0$.
One can generalize it immediately to give a spacetime
with constant curvature by adding a cosmological
constant, $\lambda^2$, to the action.
If we want to add dynamical content to the scalar field
then a term proportional to $(\partial\phi)^2$
have to be introduced in the action.
If the action is to be homogeneous in $\phi$ such
that $S[\mbox{constant}+\phi]=\mbox{constant}\times S[\phi]$,
then the proportionality term in $(\partial\phi)^2$
is a constant which we call $\omega$.
The action can then be written as
\begin{equation}
   S=\int d^2x \sqrt{-g} e^{-2\phi}
     \left( R-4\omega(\partial\phi)^2+4\lambda^2 \right).
\label{eq:5}
\end{equation}
This is the Brans-Dicke action which of course can be generalized
even more to contain a potential term of the scalar field,
$4\lambda^2\rightarrow U(\phi)$,
and to have a $\phi$-dependent Brans-Dicke parameter,
$\omega\rightarrow\omega(\phi)$ \cite{bank}.

\vskip 1cm
\noindent
{\bf 3. The Analogue of General Relativity}
\vskip 3mm

\noindent
General Relativity is the particular case of Brans-Dicke theory
when $\omega\rightarrow\infty$.
We now show that the action given in eq.~(\ref{eq:3})
can be derived in two-dimensions from the Brans-Dicke
action in the same limit.
Let us generalize action (\ref{eq:4}) to include matter,
\begin{equation}
   S=\int d^2x \sqrt{-g}
     \left\{ e^{-2\phi}
             \left[ R
                   -4\omega(\partial\phi)^2
                   +4\lambda^2 \right]
            +4{\cal L}_{matter}
     \right\},
\label{eq:7}
\end{equation}
where
$g$ is the determinant of the 2D metric,
$R$ is the curvature scalar,
$\phi$ is a scalar field,
$\lambda$ and $\omega$ are constants.
Variation of this action with respect to $g_{ab}$ and $\phi$
gives, respectively,
\begin{eqnarray}
   & &  e^{2\phi} T_{ab} =
       -2\left(\omega+1\right) D_a\phi D_b\phi
       +D_aD_b\phi
       -g_{ab} D_cD^c\phi   \nonumber \\
   & &  \ \ \ \ \ \ \ \ \ \ \ \ \ \ \ \ \ \ \ \ \
       +\left(\omega+2\right) g_{ab} D_c\phi D^c\phi
       -g_{ab}\lambda^2,  \label{eq:8}  \\
   & &  R -4\omega D_cD^c\phi
          +4\omega D_c\phi D^c\phi
          +4\lambda^2 =0,      \label{eq:9}
\end{eqnarray}
where $D$ represents the covariant derivative.
The limit $\omega\rightarrow\infty$ in eq.~(\ref{eq:9})
makes sense if one makes
\begin{equation}
    \phi=\phi_0+\frac{\varphi}{4\omega}+{\cal O}
          \left( \frac{1}{\omega^2} \right),
\label{eq:9a}
\end{equation}
where the numerical factor $\frac14$ was chosen for later
convenience.
In this case we have
\begin{equation}
   R+4\lambda^2=\Box\varphi.
\label{eq:10}
\end{equation}
The trace of eq.~(\ref{eq:8}) together with (\ref{eq:9a})
gives
\begin{equation}
   4\omega e^{2\phi_0}e^{\frac{\varphi}{2\omega}}T=
      -\Box\varphi
      +\frac{1}{2\omega}(\partial\varphi)^2
      -8\omega\lambda^2.
\label{eq:11}
\end{equation}
Then, comparing (\ref{eq:10}) and (\ref{eq:11})
we obtain
\begin{equation}
   R+4\omega e^{2\phi_0}e^{\frac{\varphi}{2\omega}}T
    +8\omega\lambda^2=
    -4\lambda^2
    +\frac{1}{2\omega} (\partial\varphi)^2.
\label{eq:12}
\end{equation}
Finally, redefining
\begin{eqnarray}
  & &  \Lambda=-8\omega\lambda^2,   \label{eq:12b} \\
  & &  \bar{T}=-4\omega e^{2\phi_0}T, \label{eq:12a}
\end{eqnarray}
and taking the limit
$\omega\rightarrow\infty$ one has
\begin{eqnarray}
  & & R=\Box\varphi, \label{eq:13} \\
  & & R-\Lambda=T,    \label{eq:14}
\end{eqnarray}
where we have dropped the bar over $T$.
Note that the limit $\omega\rightarrow -\infty$
gives the same theory.

Substituting (\ref{eq:9a})
into action (\ref{eq:7}) one obtains
\begin{equation}
S=-\frac{1}{2\omega}e^{-2\phi_0}
   \int d^2x \sqrt{-g}
   \left\{
      -2\omega R
      +\varphi R
      +\frac{(\partial\varphi)^2}{2}
      -8\omega\lambda^2
      -8\omega e^{2\phi_0} {\cal L}_{matter}
   \right\}.
\label{eq:15}
\end{equation}
In order to simplify eq.~(\ref{eq:15})
we perform the following three steps:
(i)   drop out the term $-2\omega R$ inside curly brackets,
      since variation with respect to
      $\varphi$ and $g_{ab}$ yields zero;
(ii)  renormalize the cosmological constant as in
      eq.~(\ref{eq:12b}) and the matter Lagrangean using
      eq.~(\ref{eq:12a}); and
(iii) absorve into the action
      the constant spurious coeficients in front of the integral.
The new effective action is then
\begin{equation}
   S=\int d^2x \sqrt{-g}
     \left(  \varphi R
            +\frac12 (\partial \varphi)^2
            +\Lambda
            +2{\cal L}_{matter} \right),
\label{eq:16}
\end{equation}
which is precisely equal to eq.~(\ref{eq:3}).
Since in four dimensions the Einstein's Theory of Relativity
can be obtained from the Brans-Dicke theory taking the limit
$\omega\rightarrow\infty$,
it is natural to consider the theory given by action
(\ref{eq:7}) in the limit $\omega\rightarrow\infty$
(the R=T theory) as the 2D analogue of General Relativity.
Thus,
equation (\ref{eq:14}) is the 2D equivalent of Einstein
equations;
equation (\ref{eq:13}) determines the auxiliary
field $\varphi$.

\vskip 1cm
\noindent
{\bf 5. Conclusions}
\vskip 3mm
\noindent
The theory we have shown to be 2D natural analogue of
General Relativity has been explored by Mann and
colaborators.
It admits gravitational collapse \cite{mann4,mann5},
gravitational waves \cite{mann4},
particle solutions with horizons and their causal structure
\cite{mann1,mann5a},
cosmological solutions \cite{mann4,mann6},
and semiclassical approximations \cite{mann2}.

We have found in another work \cite{sa}
that this theory also admits a vacuum black
hole with two rather surprising properties:
(i) it is a black hole of constant curvature, and
(ii) it has no spacetime singularities.
Black holes of constant curvature have also been found
in three dimensions by identifying certain points of
the anti-de~Sitter spacetime \cite{bana}.
The two dimensional constant curvature black hole version
arises through a similar procedure and is analogous
to the $\omega=0$ (see eq.~(\ref{eq:4})) black hole
\cite{lemo2}.
On the other hand non-singular black holes have also
appeared in the framework of an exact solution of 2D
string theory \cite{perr}.
Thus, it seems that the non-singular character of
these black holes is a property of the 2D world.

As a final remark we would like to point out, that
whereas we have discussed here the 2D
analogue of General Relativity,
the theory given by $\omega=-\frac12$ (see eq.~(\ref{eq:4}))
is equal to vacuum planar General Relativity \cite{lemo1}.

\vskip 1cm
\noindent
{\bf Acknowledgements}

\noindent
We wish to thank Patricio Letelier for an interesting conversation
on matters related to section~2.
JPSL acknowledges grants from JNICT (Portugal) and CNPq (Brazil).
PMS acknowledges a JNICT (Portugal) grant BIC/776/92.

\end{document}